# Reduction of Trapping and Recombination in Upgraded Metallurgical Grade Silicon: Impact of Phosphorous Diffusion Gettering

N. Dasilva-Villanueva[†]*, S. Catalán-Gómez*, D. Fuertes Marrón, J.J. Torres, M. García-Corpas, C. del Cañizo

Instituto de Energía Solar, Universidad Politécnica de Madrid, ETSI Telecomunicación, Av. Complutense 30, E-28040 Madrid (Spain).

**Abstract.** Upgraded metallurgical grade (UMG) silicon (Si) has raised interest as an alternative material for solar cells due to its low cost, low environmental impact and low CAPEX. Maximum cell efficiencies at the level of those obtained from high purity poly-Si have been reported. However, a higher defect density and the compensated doping character result in UMG-based cell efficiencies varying over wider ranges in frequency distribution charts. In this report we characterize mc-Si UMG samples with different defect densities, comparing them with monocrystalline silicon (mono-Si) UMG and commercial high-performance multicrystalline silicon (mc-Si) samples, analysing the impact of carrier trapping by means of photoconductance (PC) decay measurements, and its evolution after applying a phosphorous diffusion gettering (PDG) process. When analyzing the decay time constant of the PC measurements, slow (66.8±14.3 ms) and fast (16.1±3.5 ms) traps are found in mc-Si samples, while no evidence of trapping is found in mono-UMG samples. Slow traps are effectively removed after the PDG process, while fast traps do remain. The influence of dislocations clusters and the possible role of oxygen, as revealed by Fourier-transform infrared spectroscopy (FTIR) is discussed. Finally, the improvement in minority carrier lifetime due to the PDG treatment is reported for each sample type, reaching values up to 140 µs in mc-Si samples with no slow traps neither interstitial oxygen FTIR-peaks.

[†]Corresponding author: nerea.dasilva@ies.upm.es
*These authors contributed equally to this work

# 1. Introduction

In the last years, new additions to the overall power generation capacity based on photovoltaics (PV) have surpassed the 100 GW/year level, outpacing the annual additions of any other power source technology, both renewable and non-renewable [1]. With a total installed capacity of 760 GW at the end of 2020, it is now the second renewable energy source after hydropower [2]. In many parts of the world PV is already the cheapest alternative in terms of levelized cost of electricity (LCOE), with an associated carbon footprint well below that of fossil-fuel power technologies [3]. It is clear now that PV will be the backbone of a decarbonized electric system.

However, to successfully reach Climate Agreements, mitigating thereby the effects of carbon emissions and climate change, it is estimated that a compound annual-growth rate of PV of at least 18% needs to be maintained over the next decade [4]. Such target will only be possible if PV LCOE continues its steady decline. With material cost representing approximately 13% of the total module cost [5], the LCOE decrease will benefit from innovations in technology production. This is particularly relevant in the case of high purity silicon (Si), as its production is expensive in terms of energy demand. Additionally, new Si-purification techniques could also improve the PV life-cycle assessment with a reduced emission budget. This is the main strength of upgraded metallurgical grade (UMG) Si, a material obtained via an alternative purification route resulting in lower economic and environmental cost as compared to conventional polysilicon (poly-Si) [6–8].

Nevertheless, it is known that as-grown UMG material has a higher density of defects than conventional poly-Si, which results in low bulk minority carrier lifetime values, that can be in the range of 5-10 µs [9,10]. Furthermore, Ga, P and B dopants are routinely added into UMG-Si, causing a moderate compensated character with a potential impact on both carrier trapping and recombination [11]. With the aim of reducing bulk recombination, phosphorous diffusion gettering (PDG) processes at different conditions have been applied to UMG wafers, reporting modest lifetime improvements of around 6-fold in mc-UMG Si [12,13] and 2.3-fold in monocrystalline UMG-Si [14]. Since PDG is much more efficient in terms of lifetime improvements when applied to conventional mc-Si [15], further studies on how benign its effect can be in UMG samples, despite its efficacy greatly depending on the initial material properties.

Even so, experimental research in conventional production lines has shown that cells based on UMG-Si show a negligible penalty in terms of efficiency (with values up to 20.76%) in Al-BSF and PERC architectures, as compared with standard polysilicon-based cells. [9]

For the electrical characterization of Si, minority carrier lifetime is one of the most useful accessible parameters, and a variety of electronic processes that limit the performance of related devices can be studied thereof [16]. Lifetime data can be obtained from inductively coupled photoconductance decay (PCD) measurements following flash-light illumination of the sample under study as described in [17]. Two different measurement modes are typically applied, depending on the actual magnitude of the carrier lifetime values. In the quasi-steady-state photoconductance (QSSPC) mode [17], the photoconductance of the Si wafer is recorded while a slowly varying illumination is impinging upon the sample. A steady-state analysis can be used to determine the apparent lifetime from the absolute values of the photogeneration and the photoconductance, provided the lifetime is much lower than the decay constant of the light (typically, below 200-300 µs). For samples with higher carrier lifetime values, the transient mode is recommended, in which a fast flash excites carriers, whose contribution to photoconductance (PC) is tracked upon light extinction.

A characteristic effect, related to carrier trapping, is typically observed in PCD measurements at low carrier injection levels [19]: the lifetime dramatically increases when the excess carrier density decreases [20–22]. The temporal trapping of minority carriers at defects with large ratios of electron and hole capture cross-sections leaves behind an excess of majority carriers that contributes to the apparent conductance [23,24] . In addition, this causes the PC after the light pulse to generally deviate from a single exponential decay, revealing instead a multi-exponential character with the concurrence of more than a single electronic process involved [25]. Such processes, ascribed to trapping centers, result in prolonged apparent lifetime values that can reach hundreds of microseconds or longer, as observed in different types of Si (single crystal, polysilicon, thin film) [11,25].

Since it is well known that these trapping centers may increase the recombination of the bulk [26], some works have tried to eliminate trapping centers through procedures such as PDG and laser treatments [22,27,28]. Attempts to unveil the microscopic origin of traps in Si have only been partly successful: e.g., hints have been found about a linear relationship between oxygen concentration and trap concentration, as concluded from quantitative analyses [11,29]. Moreover, it has also been reported that dislocations do play an important role in trapping [22].

In this work we report on the influence of a PDG treatment in mc-Si UMG samples with different defect densities, and compare it with the case of monocrystalline UMG samples. We focus both on the trapping effects and also on lifetime improvement due to the PDG process, analysing up to what extent the presence of dislocations and interstitial oxygen can be related to the observed results.

## 2. Experimental section

UMG material has been manufactured via the metallurgical route by FerroGlobe, and crystallized as multicrystalline ingots by FerroSolar [9]. The resulting ingots were sliced in 15x15 cm$^2$ wafers of 200±5 μm thickness with resistivity values ranging from 1 to 1.7 Ω·cm, as shown later on. Three different types of UMG-Si wafers were analyzed and processed: two types of mc-Si wafers of two different quality levels, labelled as *A-type* and *B-type* wafers and monocrystalline wafers labelled as "*mono*". A-type samples were produced following the manufacturer's standard purification route, while B-type material resulted from mixing part of the manufacturer's feedstock with contaminated material. More details on the purification process are given elsewhere [30]. *Mono* wafers are monocrystalline 2.6 Ω·cm p-type UMG Si wafers provided by the same manufacturer, from the same UMG-feedstock, but crystallized via the Czochralski method. The compensation level of the UMG material used in this study and defined as $(N_A+N_D)/(N_A-N_D)$, where $N_A$ and $N_D$ are the concentrations of p-type and n-type dopants, respectively, varies along the ingot between 2 and 4. Commercial, polysilicon-based, 16.6x16.6 cm$^2$ mc-Si wafers were used for comparison, with resistivity values ranging 1.2-1.4 Ω·cm.

Minority carrier lifetime data were obtained from bare wafers by QSSPC measurements with the Sinton Instruments WCT-120 tool, using a slow flash decay mode with a constant decay of 1.8 ms, as measured at the reference cell. For the acquisition of the lifetime curves, an optical constant of 0.7 and the Dannhäuser carrier mobility model were used. This characterization was performed before and after 0.1M iodine-ethanol (IE) passivation for the reduction of surface recombination. To perform the surface passivation, the samples are submerged in a 0.1M IE solution after an HF treatment to remove possible silicon oxide traces on the surface of the samples. Then, the photoconductance measurement is immediately performed in order to avoid alterations in the solution due to solvent evaporation. Though different levels of surface passivation are observed in the literature and the compensated nature of the material could slightly alter those, IE passivation has been proved an efficient passivating agent leading to surface recombination velocities as low as 1 cm/s.[31]

Additionally, spatially resolved lifetime characterization by means of microwave reflectivity was carried out after IE passivation using the Semilab WT-2000 instrument, in order to locate low-lifetime regions in the samples. To characterize the trapping processes, the PC raw measurements were directly analyzed, thereby avoiding the assumptions required to convert these data into lifetime values. PC values were satisfactorily fitted to a two-exponential function of the form $y=A_1 \cdot exp(-x/t_1)+A_2 \cdot exp(-x/t_2)$, as shown in Figure *1*, following the method thoroughly explained in [28]. Note that the photovoltage signal is proportional to the PC, and that when the time constant of the first exponential $t_1$ follows the light decay, the time constant of the second exponential $t_2$ can be assigned to the dominant trap.

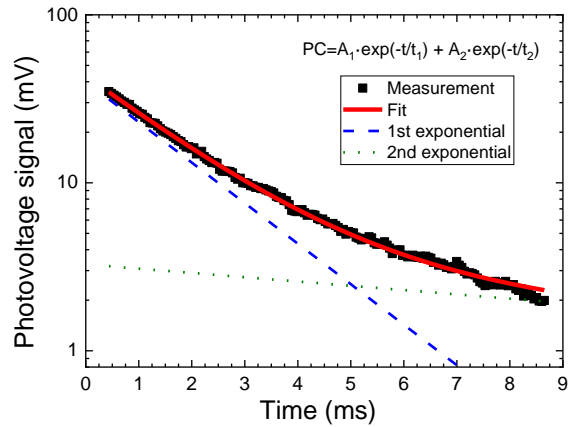

Figure 1. Fit of an exemplary photoconductance decay curve. Green dot and blue dashed lines represent the two exponential functions used to fit (solid red line) the data (black squares).

After a chemical etching with CP4 ($HNO_3$/HF) and RCA1 surface cleaning, the wafers were subjected to a thermal treatment. PDG processes were performed in a tubular furnace for different times in $O_2$ and $N_2$ atmosphere. Liquid $POCl_3$ was the P source during the process. Different temperatures were used for multi and monocrystalline samples as it was observed that the maximum improvement in lifetime occurred at a 60 minutes process at 780ºC and 820ºC, respectively. The extraction is immediate after the PDG and takes place in 5 minutes, letting the sample passively cool down. After PDG the emitter formed atop the surface was removed by CP4 etching. Then, PCD measurements were again taken without and with IE passivation and converted to lifetime curves.

All samples, both before and after PDG, were studied by means of Fourier transform infrared (FTIR) spectroscopy (Thermo Scientific Nicolet iS10) in the 400-4000 $cm^{-1}$ range in transmission mode, with the aim of detecting the fingerprint of impurities in the Si crystal. A float zone Si sample of 200 μm thickness and 3-5 Ω·cm resistivity was used as a reference sample.

After all the electrical and optical characterization, Sopori etching [32] was performed on 16 randomly selected wafers of all types in order to reveal the presence of dislocations. The wafers were then imaged with a scanner and the contrast was treated to highlight the presence of dislocations clusters.

## 3. Results and Discussion

Figure 2a represents the resistivity of the different type of as-received UMG wafers, obtained via four-point-probe measurements. Differences between mono, A-type and B-type wafers are readily visible. Mono UMG-wafers show higher resistivity values, whereas mc-UMG wafers show lower values, being B-type less resistive than A-type wafers, which stay around the same values as the conventional mc-Si wafers studied.

Lifetime values of non-passivated A-type, B-type and mono-UMG wafers are represented as a function of the excess minority carrier density (injection level) in Figure 2b. Low lifetimes are obtained in all cases (around 1 µs) and trapping, as revealed by the increased lifetime at lower injection levels, is readily visible in A-type and B-type mc-Si wafers, as well as in conventional polysilicon mc-Si wafers, but not in UMG-mono-Si wafers.

Figure 2c shows the photovoltage signal obtained under QSS conditions of A-type, B-type and mono-Si UMG wafers, as well as the photovoltage signal of the reference cell of the instrument (gray triangles). Because the high carrier recombination rate observed in the samples, the PC measurements should follow the same exponential trend of the flash; whereas that is actually the case for the UMG-mono samples, it can be clearly seen that A-type and B-type curves deviate from linearity, an effect even more notable for the latter. This deviation results from trapping, as it has been well explained in the literature [25,27].

Following the method represented in Figure *1*, the fitting of 16 A-type wafers, 16 B-type wafers and 8 mono-UMG wafers was carried out and the values obtained for the decay time constant $t_2$, assigned to the dominant trapping mechanism, are represented in Figure 2d. The fitting for 15 conventional mc-Si wafers is also included for comparison. In the case of the mono-UMG wafers (olive curves), no second exponential term is needed to fit the PCD curves, consistent with the fact that there is no evidence of trapping in those wafers. These curves can be fitted with a single exponential function with $t_1$ values around the decay time constant of the light at the reference cell (1.8 ms), indicating that the high recombination rate of the samples ensures that the PCD does follow the generation profile. On the other hand, as it can be seen in Fig. 2d, a notable difference between the values of $t_2$ is found for both types of mc-UMG material. While A-type wafers show a value of 16.1±3.5 ms (values comparable to those obtained from conventional mc-Si wafers), B-type wafers show much higher values of $t_2$, 66.8±14.3 ms. This indicates the presence of at least two types of traps dominating the trapping dynamics in the range of the first tens of ms: a fast trap present in A-type wafers, similar to the one found in conventional mc-Si ones, and a slow trap present in B-type wafers.

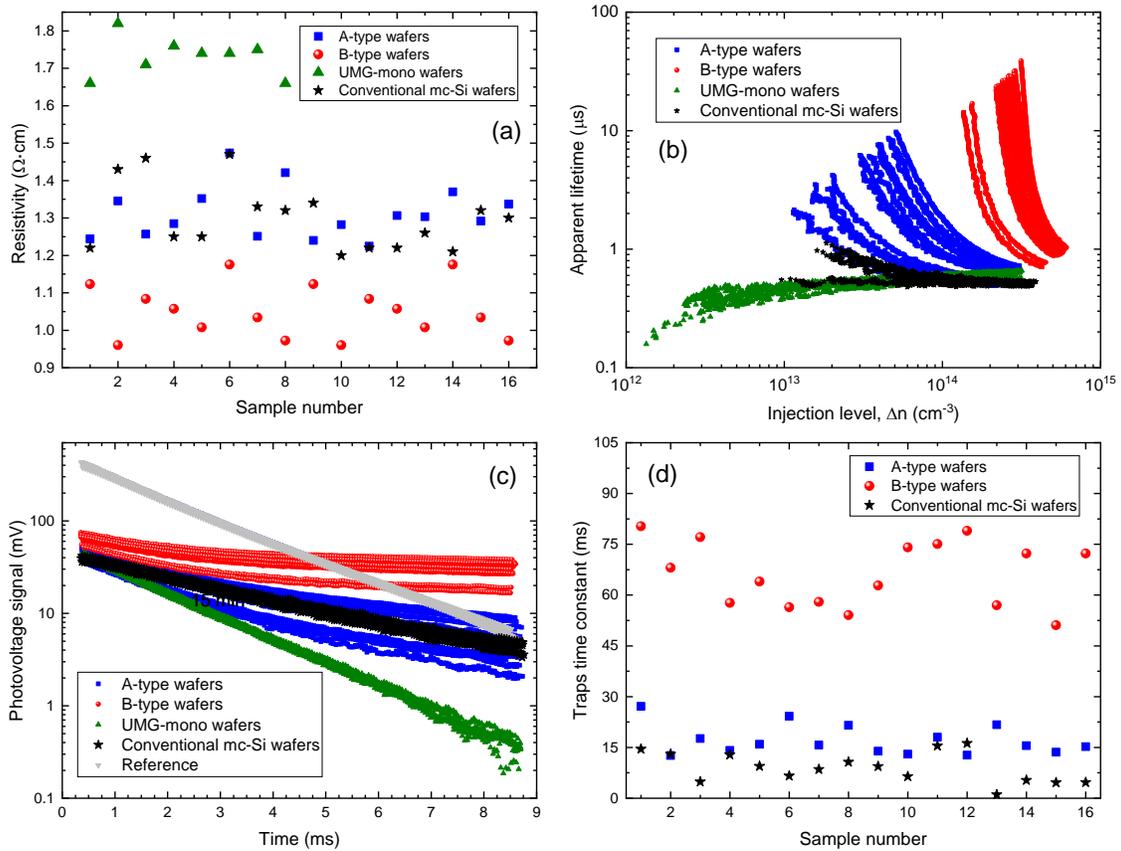

Figure 2. a) Measured resistivity values of A-type (blue squares), B-type (red spheres) and UMG-mono (green triangles) bare wafers. b) Apparent lifetime of A-type, B-type and UMG-mono bare wafers as a function of the apparent excess carrier density. c) Photovoltage signal of A-type, B-type and UMG-mono bare wafers and the output voltage of the reference cell (grey upside-down triangles) versus time. d) Trap decay time constant ($t_2$) extracted from the second exponential function of all wafers for A-type (blue squares), B-type (red spheres) and conventional multicrystalline silicon (black stars) wafers.

Figure 3a represents the PC fitting of an exemplary wafer before and after the PDG step. The first exponential, shown with the green dotted line, follows, as before, the decay of the flash, identical for both measurements, whereas the pink dashed-dotted line and the blue dashed line represent the second exponential of the fit before and after the PDG, respectively. Furthermore, as it can be seen in Figure 3a, larger values of photovoltage are obtained at short times from PDG-treated samples, indicating a reduction of the overall recombination in the sample. This result correlates well with an improvement in carrier lifetime, as shown later on. Additionally, the obtained $t_2$ values for the B-type samples after PDG are now different, as it can be seen in Figure 3a from the slope of the curve in the long-term range.

Figure 3b shows, in a box-type plot, the $t_2$ values obtained for all A-type and B-type wafers before and after the gettering. As it can be observed, similar values of $t_2$, ranging around 15 ms, are found after PDG in both types of samples. This indicates that the PDG process has been effective in removing the slow traps originally present in B-type wafers, being probably related with a getterable impurity [28]. This is however not the case for the fast trap present in A-type wafers, which is PDG-insensitive. Interestingly, $t_2$ values of B-type wafers after PDG seem to coincide with the corresponding values observed in type-A samples, pointing to a common microscopic origin in both cases.

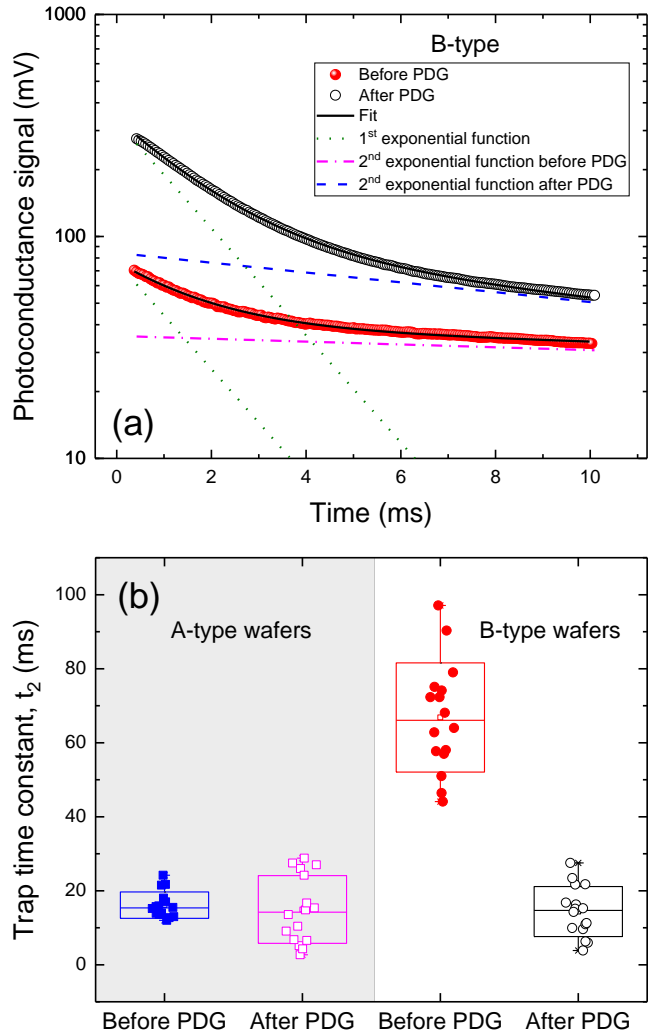

Figure 3. a) Photovoltage signal of a B-type wafer before (red spheres) and after (white circles) PDG. Green dotted lines, blue dashed line and pink dot-dashed line represent the exponential functions used to fit (solid black line) an exemplary curve. b) Comparison of $t_2$ values obtained from the fitting of the PC measurements before and after PDG for A-type and B-type wafers. In the box-type plot, whiskers represent minimum and maximum values, whereas the inside line and the height of the box represent the mean and the standard deviation, respectively.

The origin of slow traps in B-type UMG-Si is unclear. Its eventual relationship to the presence of dislocations and interstitial oxygen has been studied, as explained in the following lines.

Dislocations, among other structural crystallographic defects, can contribute significantly to minority carrier recombination, lowering thereby the overall carrier lifetime of the wafers [33,34]. This is specially the case if they are decorated by transition metal impurities [35], which is an important factor to take into account in a material with high impurity content, such as UMG-Si.

Figure 4 shows the uneven distribution of dislocations observed among samples of different types as revealed by Sopori processing of the samples. Three exemplary samples are shown in Fig. 4: one representative B-type sample, and two A-type samples from the same original wafer but from different positions, in which different values of carrier lifetime after PDG had been measured. Such differences in the observed lifetime of passivated A-type samples, that will be explained later, can reach up to an order of magnitude, following a pattern among neighboring wafers (wafers from the same part of the ingot, presenting the same grain structure).

Slight differences in grain boundaries can be perceived between the A-type samples, with the high lifetime sample on the left and the low lifetime sample on the right, and with the second one showing slightly larger grains in the central part. As it can be observed in Figure 4b, a conspicuous dislocation cluster is present in the A-type sample with low lifetime. It can also be seen how neither the high-lifetime A-type wafer nor the B-type wafer show evidence of the presence of dislocations. This procedure was carried out on eight different samples of each type, with similar results. It is therefore concluded that the slow traps observed in B-type wafers cannot be directly associated to the presence of dislocations (in the case of poly-based mc-Si wafers this has been indeed the case [21]). Sopori etching was also carried out on mono-Si UMG wafers, which, as expected, did not reveal the presence of dislocations.

Lifetime maps obtained from µW-PCD measurements at low-injection levels, included in Figure 4c, show that the lower average lifetime of the "bad" A-type wafer is due to a low-lifetime region in the center of the sample, which coincides with the region where dislocations are found. When represented in the same color scale, B-type wafers show an overall and homogenous lower lifetime than A-type wafers. It is therefore concluded that dislocation clusters unevenly distributed among wafers are presumably responsible of the wide spread of carrier lifetime values observed in UMG-Si.

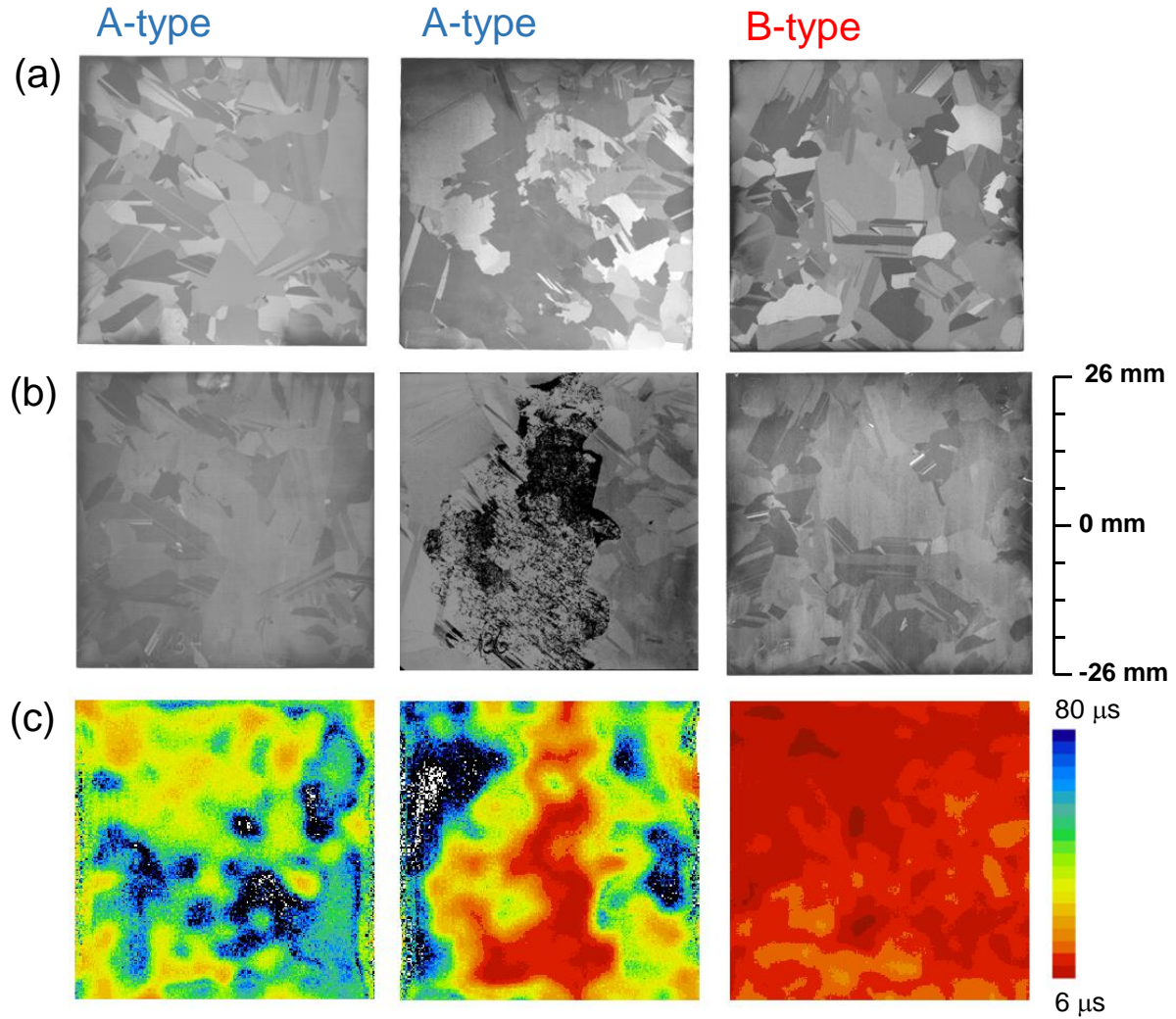

Figure 4. a) Images of two A-type wafers and one B-type wafer after PDG, before the Sopori etching. b) Dislocations maps of the same two A-type wafers and the B-type wafer after Sopori etching, enhanced via image-processing software (dark contrast is localized in regions with a high dislocation density). c) Lifetime scans of the same two A-type wafers and the B-type wafer after PDG, before the Sopori etching, with IE surface passivation.

Figure 5 shows FTIR spectra of the samples under study. The transmittance is plotted as a function of the wavenumber for the three types of wafers before (a) and after (b) the PDG process. The spectra are represented in the range from 900 to 1300 cm$^{-1}$ where the oxygen-related bands are present [36]. Specifically, the band at 1107 cm$^{-1}$ is ascribed to antisymmetric stretching of interstitial oxygen atoms in the quasi-molecule Si-O$_i$-Si, and the band around 1230 cm$^{-1}$ corresponds to platelet precipitates [37].

Two bands at 1107 and 1230 cm$^{-1}$ can be observed both in the mono and B-type wafers, while only the latter can be discerned in the A-type wafers. Additional bands reported at 1060 and 1180 cm$^{-1}$ due to other precipitate shapes have not been detected [38,39]. The presence of silicon dioxide precipitates (band at 1230 cm$^{-1}$) has been previously reported not only in conventional Si [40,41] but also in UMG material [13]. Interestingly, A-type samples are the only ones that do not show evidence of $O_i$ at the bare wafer stage. The disappearance of the band at 1230 cm$^{-1}$ after PDG can be clearly observed in all samples in Figure 5b. Although the dissolving of the precipitates is typically accompanied with an increment of the $O_i$ band [36] (at 1107 cm$^{-1}$), apparent changes in the latter band have not been found. In fact, this peak is only observed in the samples which already had the $O_i$ band before the PDG (mono and B-type samples), indicating the inefficacy of PDG at the proposed temperatures in removing $O_i$. Additionally, it can be concluded that there seems to be no direct link between the presence of $O_i$ and the origin of the slow trap observed in B-type wafers, since the latter is removed after PDG (Figure 3).

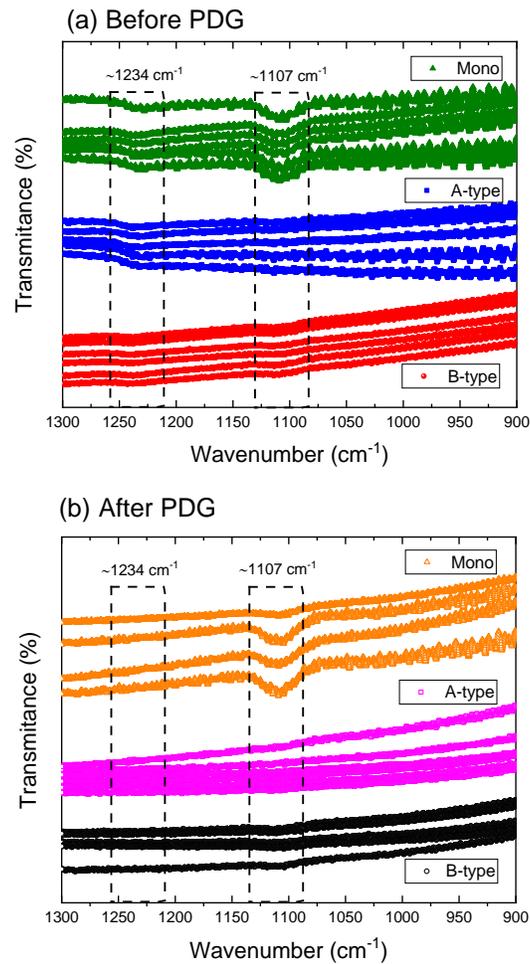

Figure 5. Transmittance vs wavenumber, obtained from FTIR analysis, of mono-UMG, A-type and B-type wafers a) before the PDG and b) after the PDG. Bands at 1107 cm$^{-1}$ and 1230 cm$^{-1}$ ascribed to interstitial (Oi) oxygen and oxygen precipitates, respectively, are highlighted in the figure.

Figure 6a and Figure 6b show the passivated lifetime curves for the mono-UMG wafers and the mc-Si wafers (both A-type and B-type), respectively, before and after the PDG process. Figure 6c is a comparative plot of the lifetime values for all types of wafers before and after the PDG, evaluated at an injection level of $10^{15}$ cm$^{-3}$.

The carrier lifetime of mono wafers does not improve upon the same PDG conditions used for UMG wafers. However, improvement was found for the PDG performed at higher temperatures, 820ºC, as shown in Figure 6c, where the improvement in UMG-multi wafers was reduced.

For mc-Si UMG wafers, a considerable improvement, both in lifetime values and injection level range, is perceived. However, it is important to note that even though A- and B-type wafers show similar passivated lifetime values before the PDG, A-type wafers show a much higher improvement after the PDG process, almost doubling the improvement shown by B-type wafers, and surpassing 100 µs.

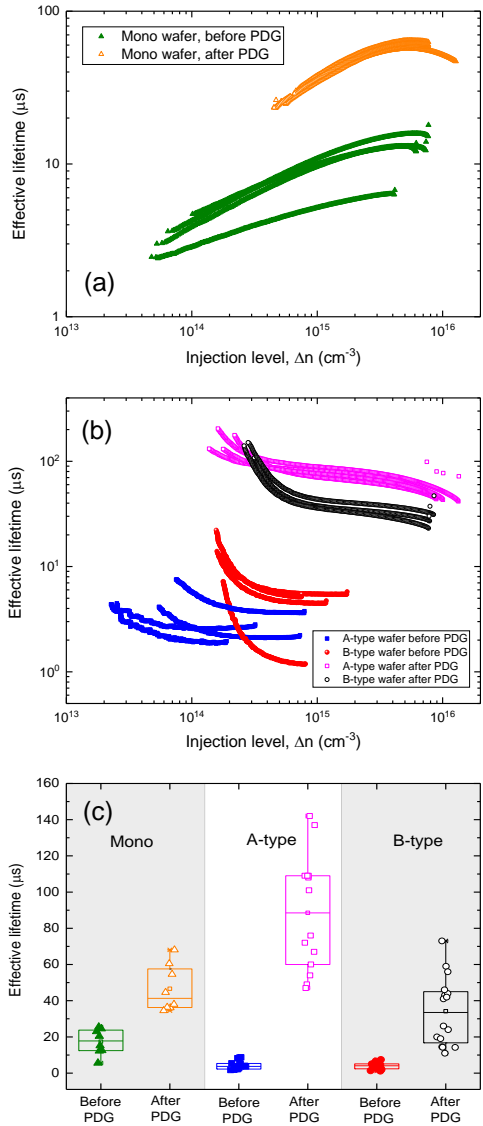

Figure 6. a) Carrier lifetime of eight passivated mono-UMG wafers as a function of the injection level before (green triangles) and after (orange triangles) the PDG process at 820ºC. b) Lifetime of four passivated A-type and four B-type wafers as a function of the injection level before and after the PDG process at 780ºC. c) Lifetime values of passivated samples evaluated at an injection level of $10^{15}$ cm$^{-3}$ of mono-UMG, A-type and B-type wafers before and after the PDG process (820ºC/780ºC). In the box-type plot, whiskers represent minimum and maximum values, whereas the inside line and the height of the box represent the mean and the standard deviation respectively.

The comparison of the FTIR and lifetime data before and after PDG seems to suggest that the cause of the worse PDG performance on the B-type samples (with medium $O_i$ content and modest lifetime improvement) and mono wafers (with a high $O_i$ and reduced lifetime improvement compared to A-type wafers) could be related to the presence of $O_i$, considering in addition that the samples with the highest lifetime improvement upon PDG (A-type) do not show the presence of $O_i$ prior to the treatment. This observation is in agreement with the results of Sabatino *et al* [42], who demonstrated that a high density of old thermal donors makes the lifetime in UMG material to decrease after PDG, and with the results of Binetti *et al* [13], who reported the detrimental role of $O_i$ in UMG-Si and the inefficacy of the PDG when it is present in high concentration. The presence of oxide precipitates has been also previously linked to lower PDG improvements under high impurities concentrations [43]; which could explain the reduced lifetime improvement in B-type wafers, given the impurity concentration is higher when compared to A-type wafers.

However, we cannot correlate the slow trapping centers in this UMG material with the presence of thermal donors since $O_i$ is still present in B-type wafers after PDG (Fig 5 b), while the slow traps have been removed after PDG (Fig. 3 b) for all these wafers. Furthermore, mono wafers show the $O_i$ peak before and after PDG but no trapping in any case.

## 4. Conclusions

In summary, trapping in UMG mc-Si wafers can be accessed and analyzed by QSS photoconductance decay measurements without the need of translating data into carrier lifetime values. This procedure is advantageous for low-lifetime samples, since it eliminates eventual inaccuracies due, e.g., to the carrier mobility model of choice for a compensated material. PC measurements could be satisfactorily fitted to a double-exponential decay curve, the first term accounting for carrier recombination following the flash decay (an important condition, fulfilled by low-lifetime samples), and the second term ascribed to the presence of carrier traps. Two types of traps, fast and slow, have been identified in A- and B-type mc-Si samples, respectively, while no trapping effects could be observed in UMG mono-Si wafers.

Though slow traps are only present in B-type mc-UMG wafers, the lifetimes recorded from passivated mc-UMG wafers of all types are low, 5 µs in average at $10^{15}$ cm$^{-3}$, indicating that the type of trap, fast or slow, is actually not limiting bulk lifetime values. Mono-UMG wafers show much higher lifetimes at bare wafer stage, up to one order of magnitude higher that mc-UMG wafers. Still, lifetime values of UMG material of all sorts need to be improved for ensuring a good performance in terms of solar cell efficiency. The analysis of PC measurements shows that after a PDG process slow traps are removed from B-type wafers, but fast traps, likely of the same type present in A-type samples, still remain. This result indicates that slow traps are getterable, together with recombination centers. However, no change was observed in the decay time constant for A-type wafers.

Sopori etching followed by image digitalization proved that clusters of high dislocation densities were present in some A-type samples, but not in B-type wafers, a result that rules out dislocations as the origin of slow traps. FTIR analysis of UMG wafers showed that a peak at 1107 cm$^{-1}$, ascribed to interstitial oxygen and present only in B-type and mono wafers is not modified after PDG while oxygen precipitates (peak around 1234 cm$^{-1}$) disappear from all types of UMG wafers.

Through a PDG process, lifetime values in mc-UMG wafers could be significantly improved, reaching 150 us for A-type wafers and representing an improvement of up to 25-fold. This improvement is less significant for B-type wafers, where slow traps and $O_i$ were present before the PDG. Slow trapping centers are removed from B-type wafers after PDG, while $O_i$ still remains.

**Acknowledgements**

The Spanish Agencia Estatal de Investigación is acknowledged for funding through the SOLAR-ERA.NET Cofund project "Low Cost High Efficient and Reliable UMG PV cells (CHEER-UP)", PCI2019-111834-2 /AEI/10.13039/501100011033. The Comunidad de Madrid is also acknowledged for support under the project Madrid-PV2 (S2018/EMT-4308). S.C.G acknowledges Juan de la Cierva en Formación programme (reference FJC2019-041616-I). Bo-Kyung Hong and Manuel Funes are acknowledged for support in wafer processing and for fruitful discussions. Aurinka PV is acknowledged for UMG wafer supply.